\documentstyle[12pt]{article}
\newcommand{\hs}{\hspace{.15cm}}
\def\mm#1{ m_{\tilde #1}^2  }
\def\smm#1{   m_{ #1}^2   }
\def\s#1{   \theta_{ #1}  }
\def\ss#1#2{\theta_{{ #1}{#2}}    }
\def\t#1{   \theta_{\tilde #1}  }
\def\st#1#2{   \theta_{{ #1}{\tilde #2}}   }
\def\tt#1#2{   \theta_{{\tilde #1}{\tilde #2}}   }
\def\tt#1#2{   \theta_{{\tilde #1}{\tilde #2}}   }

\def\ttt#1#2#3{   \theta_{{\tilde #1}{\tilde #2}{\tilde #3}}   }
\begin{document}
\begin{titlepage}
\begin{centering}
{\large{\bf Radiative Electroweak symmetry breaking
 in the MSSM and Low Energy Thresholds}}\\
\vspace{.5in}
{\bf A. Dedes  }\\
\vspace{.2in}
Division of Theoretical Physics,
University of Ioannina
\\
Ioannina, GR - 451 10,
GREECE\\
\vspace{.3in}
{\bf A. B. Lahanas }  \\
\vspace{.2in}%
University of Athens, Physics Department,
Nuclear
and Particle Physics Section\\
\vspace{.05in}
Ilissia, GR - 157 71  Athens,
GREECE\\
\vspace{.2in}
and
\vspace{.2in}\\
{\bf K. Tamvakis}   \\
\vspace{.2in}
Division of Theoretical Physics,
University of Ioannina
\\
Ioannina, GR - 451 10,
GREECE\\
\vspace{.4in}
{\bf Abstract}\\
\vspace{.1in}
\end{centering}
{\noindent
We study Radiative Electroweak Symmetry Breaking in the Minimal Supersymmetric
 Standard Model (MSSM).
We employ the 2-loop Renormalization Group equations for running masses
 and couplings taking into
account sparticle threshold effects. The decoupling of each
particle below its threshold is realized by  a step function in all one-loop
Renormalization Group equations (RGE). This program
requires the calculation of all wavefunction, vertex  and mass
renormalizations for all particles involved.
Adapting our numerical routines to take care of the succesive decoupling of
each particle below its threshold, we compute the mass spectrum of sparticles
and Higgses consistent with the existing experimental constraints.
The effect of the threshold corrections is in general of the same order of
magnitude  as the two-loop contributions with the exception of the
heavy Higgses and those neutralino and chargino states that are nearly
Higgsinos for large values of the parameter $\mu$.}

\par
\vspace{4mm}
\begin{flushleft}
IOA-315/95 \\
UA/NPPS -
17/1995\\
\end{flushleft}
\end{titlepage}

\baselineskip=24pt
\par
\noindent
{\bf 1. Introduction}
\\

The low energy values of the three gauge coupling constants
known to the present experimental accuracy rule out the simplest
 versions
of the Grand Unified Theories. In contrast, supersymmetric unification,
 in the
framework of Supersymmetric Grand Unified
Theories$^{\cite{Nilles}}$ $^{\cite{witten}}$, is in
excellent agreement$^{\cite{amaldi}}$
 with a unification energy
scale $M_{GUT}$ within the proton decay lower bounds. Moreover,
 softly broken supersymmetry,
possibly
resulting from an underlying Superstring framework, could lead
 to $SU(2)_{L}\times U(1)_{Y}$ gauge
symmetry breaking through radiative corrections for a certain range
 of values of the existing free
parameters$^{\cite{Iban}}$. In such a scenario the elegant ideas of
 Supersymmetry, Unification and
Radiative Symmetry Breaking are realized within the same framework.
 The Minimal
Supersymmetric extension of the Standard Model (MSSM)$^{\cite{Nilles}}$
incorporates all the
above. Due to its minimal content and the radiatively induced symmetry
 breaking
it is the most predictive of analogous theories.

As in the numerous$^{\cite{Ross}\cite{castano}}$ existing analyses of
 radiative symmetry breaking in
the MSSM, in the present article we employ the Renormalization Group.
The Higgs boson running mass-squared matrix, although positive
definite at large energy scales of the order of $M_{GUT}$, yields a
negative eigenvalue at low energies causing the spontaneous
breakdown of
the electroweak symmetry. The ``running" of mass parameters from
 large to low
scales is equivalent to computing leading logarithmic radiative
 corrections.
Although this scenario depends on the values  of few (3 or 4)
free parameters,
one could interpret it as leading to the prediction of $M_{Z}$
in terms of
$M_{GUT}$, or the Planck mass, and the top quark Yukawa coupling.
 Another way to
interpret the predictions of this model is to consider $M_{Z}$
 determined in
terms of the supersymmetry breaking scale. The analysis of the
 results helps us
find out to which extent the low energy data can constrain the
type and scale
of supersymmetry breaking.

The purpose of the present article is to include in the above stated
 scenario
the so-called low-energy ``threshold effects". Since we have employed
the
${\overline{DR}}$ scheme in writting down the one-loop Renormalization
 Group
 equations, which is by definition mass-indepedent, we could ``run"
them from
$M_{GUT}$ down to $M_{Z}$ without taking notice of the numerous
sparticle
thresholds existing in the neighborhood of the supersymmetry breaking
 scale
near and above $M_{Z}$. This approach of working in the ``full"
theory
consisting of particles with masses varying over 1-2 orders of
 magnitude has
to overcome the technical problems of the determination of the pole
 masses. Our
approach, also shared by other analyses, is to introduce a succession
 of
effective theories defined as the theories resulting after we
functionally
integrate out all heavy degrees of freedom at each particle threshold.
 Above
and below each physical threshold we write down the Renormalization
Group
equations in the ${\overline{DR}}$ scheme only with the degrees of
freedom
that are light in each case. This is realized by the use of a theta
function at
each physical threshold. The integration of the Renormalization Group
equations
in the ``step approximation" keeps the logarithms $\ln(\frac{m}{\mu})$
 and
neglects constant terms. The physical masses are determined by the
condition
$m(m_{phys})=m_{phys}$ which coincides with the pole condition if we
keep
leading logarithms and neglect constant terms. The great advantage of
 this
approach is that the last step of determining the physical mass
 presents no
extra technical problem and it is trivially incorporated in the
integration of
the Renormalization Group equations.
\\
\\
\\
{\bf 2.  The softly broken Minimal Supersymmetric Standard Model}
\\

The superpotential of the minimal supersymmetric extension of
the standard
model, or just MSSM, is
\begin{equation}
{\cal W}=Y_{e}L^{j}E^{c}H_{1} ^{i}{\epsilon_{ij}}
	 + Y_{d}Q^{ja}D_{a} ^{c}H_{1} ^{i}{\epsilon_{ij}}
	 + Y_{u}Q^{ja}U_{a} ^{c}H_{2} ^{i}{\epsilon_{ij}}
	 + \mu H_{1} ^{i}H_{2} ^{j}{\epsilon_{ij}}
\end{equation}
$(\epsilon_{12}=+1)$ in terms of the quark $Q(3,2,1/6)$,
 $D^{c}(\overline
3,1,1/3)$,  $U^{c}(\overline 3,1,-2/3)$, lepton\newline
 $L(1,2,-1/2)$,
$E^{c}(1,1,1)$ and Higgs $H_{1}(1,2,-1/2)$, $H_{2}(1,2,1/2)$
chiral
superfields. We have suppressed family indices. The second Higgs
doublet $H_{2}$
is necessary in order to give mass to the up quarks since
the conjugate of
$H_{1}$ cannot be used due to the analyticity of the superpotential.
It is also
required in order to cancel the new anomalies generated by the
fermions in
$H_{1}$. Note that the superpotential (1) is not the most general
$SU(3)_{C}\times SU(2)_{L}\times U(1)_{Y}$ -invariant superpotential
 that can be
written in terms of the given chiral superfields since terms like
$D^{c}D^{c}U^{c}$,$QLD^{c}$,...etc not containing any ordinary standard
 model
interaction could be present. The superpotential (1) could be arrived
 at by a
straightforword supersymmetrization of the standard three Yukawa
interaction
terms. It possesses an anomalous R-parity broken by
supersymmetry-breaking
gaugino masses down to a discrete R-parity which ascribes -1 to
 matter and +1 to
Higgses. There is also an unwanted continuous PQ-type symmetry
 leading to an
observable electroweak axion which is broken by the last term in (1).
 This term
introduces a scale $\mu$ which has to be of the order of the soft
supersymmetry
breaking scale in order to achieve electroweak breaking at the observed
$M_{Z}$
value. Although it appears unnatural that the scale of breaking of
a PQ symmetry
should be related to the supersymmetry breaking, there exist schemes
 based on an
enlarged framework (extra fields or non-minimal supergravitational
couplings)
that lead to dynamical explanation of the order of magnitude of the
scale $\mu$$^{\cite{kim}}$.

The fact that supersymmetry is not observed at low energies requires
the
introduction of extra supersymmetry breaking interactions.
This is achieved by
adding to the Lagrangian density, defined by the given
$SU(3)_{C}\times SU(2)_{L}\times U(1)_{Y}$ gauge symmetry and $\cal W$,
extra
interaction terms that respect the gauge symmetry but break
 supersymmetry. This
breaking however should be such that no quadratic divergences
appear and the
technical ``solution" to the hierarchy problem is not spoiled.
Such terms are
generally termed ``soft". The most general supersymmetry breaking
interaction
Lagrangian resulting from spontaneously broken Supergravity in the
flat limit
$(M_{P}{\rightarrow}{\infty})$ contains just four types of soft terms ,
 i.e
gaugino masses, ${\Phi^{\ast}\Phi}$-scalar masses,
$\Phi\Phi\Phi$-scalar cubic
superpotential interactions and $\Phi\Phi$-scalar quadratic
superpotential
interactions. For the MSSM this amounts to
\begin{eqnarray}
{\cal L}_{{SB}} &=&
- {\frac {1} {2}} {\sum_{A}} M_{A} {\bar{\lambda}_{A}}{\lambda_{A}}
	   -m_{H_{1}}^{2} |H_{1}|^{2}
	   -m_{H_{2}}^{2} |H_{2}|^2
				-{m^{2} _{\tilde{Q}}}  |\tilde{Q}|^2
	   -{m^{2}_{\tilde{D}}} |\tilde{D^{c}}|^{2}
	   - {m^{2}_{\tilde{U}}} |{\tilde{U}}^{c}|^{2}\nonumber\\
		&  -& {m^{2}_{\tilde{L}}} |\tilde{L}|^{2}
	   -{m^{2}_{\tilde{E}}} |\tilde{E^{c}}|^{2}
	   - (Y_{e} A_{e} \tilde{L}^{j} \tilde{E}^{c} H_{1} ^{i}
\epsilon_{ij}
	   +Y_{d} A_{d} {\tilde{Q}}^{ja} {\tilde{D}_{a}} ^{c} H_{1} ^{i}
\epsilon_{ij}\nonumber\\
      & +& Y_{u} A_{u} {\tilde{Q}}^{ja} {\tilde{U}_{a}} ^{c} H_{2} ^{i}
\epsilon_{ij}
       +h.c)-
		(B \mu H_{1} ^{i} H_{2} ^{j} \epsilon_{ij}  +h.c)
\end{eqnarray}

Again we have suppressed family indices. We denote with $H_{1}$,$H_{2}$
the
ordinary Higgs boson doublets and with
$\tilde{Q}$,$\tilde{D^{c}}$,$\tilde{U^{c}}$,$\tilde{L}$,$\tilde{E^{c}}$
 the
squark and slepton scalar fields .The gauginos $\lambda_{A}$ are
considered as
four component Majorana spinors. Apart from the three gaugino masses,
 $B$ and the
two soft Higgs masses, there are also $5 N_{G}$ masses and $3 N_{G}$
dimensionfull cubic couplings in the simplest case that we retain only
 family
diagonal couplings.Thus, totally $6+ 8 N_{G}$
 new
parameters. Note that these new parameters are dimensionfull and that
 without a
simplifying principle they could in general represent different scales.

A dramatic simplification of the structure of the supersymmetry
 breaking
interactions is provided either by Grand Unification assumptions or
 by
Superstrings. For example, $SU(5)$ unification implies at tree
 level
$m_{\tilde{Q}}=m_{\tilde{U}}=m_{\tilde{E}}$,
$m_{\tilde{L}}=m_{\tilde{D}}$ ,
$M_{1}=M_{2}=M_{3}$ and $A_{d}=A_{e}$. $SO(10)$ unification implies
further
equality of all sparticle masses, equality of Higgs masses and
 equality of
the three types of cubic couplings. The simplest possible choice
at tree level
is to take all sparticle and Higgs masses equal to a common mass
 parameter
$m_{o}$, all gaugino masses equal to some parameter $m_{1/2}$ and
all cubic
couplings flavour blind and equal to $A_{o}$. This situation is
 common in the
effective Supergravity theories resulting from Superstrings but there
 exist
more complicated alternatives. For example Superstrings with massless
 string
modes of different modular weights lead to different sparticle masses
at tree
level$^{\cite{matalliotakis}}$. The equality of gaugino masses can also
 be circumvented in an
effective supergravity theory with a suitable non-minimal gauge kinetic
 term$^{\cite{ellis}}$. Note
however that such non-minimal alternatives like flavour dependent
sparticle
masses are constrained by limits on FCNC processes. In what follows
 we shall
consider this simplest case of four parameters $m_{o}$, $m_{1/2}$,
 $A_{o}$ and
$B_{o}$.
\\
\\
\\
{\bf 3.  Radiative corrections and symmetry breaking}
\\

The scalar potential of the model is a sum of three terms, being the
contribution of F-terms,
\begin{eqnarray}
{V_{F}} &=& |{\sum} (Y_{d} \tilde{Q}^{ja} \tilde{D}_{a} ^{c}
      +Y_{e} \tilde{L}^{j} \tilde{E}^{c}) \epsilon_{ij}
      +\mu H_{2} ^{j} \epsilon_{ij}|^{2}
      +{\sum} |Y_{e} H_{1} ^{j} \tilde{E}^{c}
\epsilon_{ij}|^{2}\nonumber \\
     &+&|{\sum} (Y_{u} \tilde{Q}^{ja}\tilde{U}_{a} ^{c}) \epsilon_{ij}
      -\mu H_{1} ^{j} \epsilon_{ij}|^{2}
      +{\sum} |Y_{e} H_{1} ^{i} \tilde{L^{j}}
\epsilon_{ij}|^{2}\nonumber \\
     &+&{\sum} |Y_{d} H_{1} ^{j} \epsilon_{ji} \tilde{D}_{a} ^{c}
      +Y_{u} H_{2} ^{j} \epsilon_{ji} \tilde{U}_{a} ^{c}|^{2}
      +{\sum} |Y_{d} H_{1} ^{i} \epsilon_{ij} \tilde{Q}^{ja}|^{2}
      +{\sum} |Y_{u} H_{2} ^{i} \tilde{Q}^{ja} \epsilon_{ij}|^{2}
\end{eqnarray}
where the sums are over the omitted family indices, the contribution
 of the
D-terms 
\begin{eqnarray}
V_{D}&=&{\frac {1} {2}} g'^2\left[-{\frac {1} {2}} |H_1|^2
      +{\frac {1} {2}} |H_2|^2
      +\sum (-{\frac {1} {2}} |\tilde{L}|^2
      +|\tilde{E^{c}}|^{2}
      -{\frac {2} {3}} |\tilde{U^{c}}|^{2}
      +{\frac {1} {3}} |\tilde{D^{c}}|^{2}
+{\frac {1} {6}} |\tilde{Q}|^{2})\right]^{2}\nonumber\\
     &+& {\frac {1} {2}} g_{3} ^{2}\left[ \sum(\tilde{Q^{\dagger}}
{\frac
{\lambda^{A}} {2}} \tilde{Q}
-\tilde{D^{c}}{\frac {\lambda^{A}} {2}} \tilde{D^{c}} ^{\dagger}
-\tilde{U^{c}} {\frac {\lambda^{A}} {2}}
\tilde{U^{c}}^{\dagger})\right]^{2}\nonumber\\
     &+& {\frac {1} {8}} g^{2} \left[|H_{1}|^{4}+|H_{2}|^{4} + (\sum
|\tilde{Q}|^{2})^{2} +(\sum |\tilde{L}|^{2})^{2}
-2 |H_{1}|^{2} |H_{2}|^{2}\right.\nonumber\\
    &-& 2 |H_{1}|^{2} \sum |\tilde{Q}|^{2} -2
|H_{2}|^{2} \sum |\tilde{Q}|^{2} - 2 |H_{1}|^{2} \sum |\tilde{L}|^{2} -2
|H_{2}|^{2} \sum |\tilde{L}|^{2}\nonumber\\&-&2 \sum |\tilde{Q}|^{2}
\sum
|\tilde{L}|^{2}+4 |H_{1} ^{\dagger} H_{2}|^{2}+4 \sum \tilde{Q_{i}
 ^{\dagger}}\tilde{Q_{j}} \sum \tilde{L_{j} ^{\dagger}} \tilde{L_{i}}
   +4 \sum |H_{1} ^{\dagger}
\tilde{L}|^{2}\nonumber\\& +&\left.4 \sum |H_{2} ^{\dagger}
\tilde{L}|^{2}+4
\sum |H_{1} ^{\dagger} \tilde{Q}|^{2}+4 \sum |H_{2} ^{\dagger}
\tilde{Q}|^{2}\right]
\end{eqnarray}
and finally the scalar part of $-{\cal{L_{SB}}}$ shown in (2).
The tree level
scalar potential  leads to electroweak breaking as long as
\begin{eqnarray}
m_{1} ^{2} m_{2} ^{2} - {\mu^2} {B^2} < 0 \nonumber
\end{eqnarray}
This is clear from
\begin{eqnarray}
V &=& m_{1} ^{2} |H_{1}|^{2} +m_{2} ^{2} |H_{2}|^{2} +
(B \mu H_{1} ^{i} H_{2} ^{j} \epsilon_{ij}  +h.c)   \nonumber \\
&+& {\frac {1} {8}} g'^{2} (|H_{1}|^{2} -|H_{2}|^{2})^{2} +
{\frac {1} {8}} g^{2} (|H_{1}|^{4} +|H_{2}|^{4} +
4 |H_{1} ^{\dagger} H_{2}|^{2}
-2 |H_{1}|^{2} |H_{2}|^{2}) +.....
\end{eqnarray}
written in terms of
\begin{equation}
   m_{1,2} ^{2} \equiv m_{H_{1,2}} ^{2} + {\mu}^{2}  .
\end{equation}

We have replaced the
appearing parameters with their running values
$m_{1} ^{2} (Q)$, $m_{2} ^{2}
(Q)$,... as defined by the Renormalization Group.
The results based on the tree level study of the potential cannot
be always trusted since they are sensitive to the choice of the
renormalization scale.
 Adding
the one-loop
radiative corrections obtained in the $\overline{DR}$ scheme,
\begin{equation}
\Delta V_{1} = {\frac {1} {64 { \pi}^{2}}} Str\{{\cal M}^{4} (ln(
{\cal M}^{2}/ {Q^{2}})-3/2)\} \end{equation}
we end up with an Effective Potential that upon minimization
supports a vacuum
with spontaneously broken electroweak
symmetry$^{\cite{castano}\cite{zwirner}}$.
A reasonable
approximation to (7) would be to allow only for the
dominant top-stop loops. Note that although the
Renormalization Group improved tree level potential
depends on the scale Q
this is not the case for the full 1-loop Effective
Potential which is
Q-independent up to,  irrelevant for minimization, Q-dependent but
field-independent terms.

Minimization of the 1-loop Effective Potential gives two conditions
\begin{equation}
{\frac {1} {2}} M_{Z} ^{2} = {\frac {{\overline m_{1} ^{2}} -
{\overline
m_{2} ^{2}} tan^{2}{\beta}} {tan^{2}{\beta}  -1}}
\end{equation}
and
\begin{equation}
sin2{\beta} =- {\frac {2 B \mu} {{\overline m_{1} ^{2}} +
{\overline m_{2}
 ^{2}}}}
\end{equation}
The angle $\beta$ is defined as
$\beta = tan^{-1} (\upsilon_{2}/\upsilon_{1})$
in terms of the Higgs v.e.v.'s
\newline $\upsilon_{1,2} = <H_{1,2} ^{o}>$. The
masses appearing in (8) and (9) are defined as
\begin{equation}
{\overline m_{1,2} ^{2}} \equiv m_{1,2} ^{2} +
{\frac {\partial (\Delta
V_{1})} {\partial \upsilon_{1,2} ^{2}}}
\end{equation}
All parameters are Q-dependent. At $Q = M_{Z}$
\begin{equation}
M_{Z} ^{2} = {\frac {1} {2}}(g^{2} + g'^{2})(\upsilon_{1} ^{2} +
\upsilon_{2}^{2}) + ...
\end{equation}
  $M_{Z}$
denotes the Z - boson pole mass $M_{Z} = 91.187 GeV$, and the ellipses
are higher order corrections.
Note also that in our
convention $\mu B$ has the same sign with ${\overline m_{1} ^{2}}+
{\overline m_{2} ^{2}}$.

We shall assume that at a very high energy scale $M_{GUT}$ the soft
supersymmetry
breaking is represented by four parameters $m_{o}$, $m_{1/2}$,
$A_{o}$ and $B$
of which we shall consider as input parameters only the first
three and treat
$B(M_{Z})$ as determined through equation (9). Actually we can treat
$\beta(M_{Z})$ as input parameter and both $B(M_{Z}), \mu(M_Z)$
are determined by solving the minimization conditions (8) and (9),
with the sign of $\mu$ left undetermined.
The top-quark mass$^{\cite{abe}}$, or
equivalently the top-quark Yukawa coupling, although localized
 in a small range
of values should also be considered as an input parameter since
the sparticle spectrum and the occurrance of symmetry breaking
itself is
sensitive to its value.Thus, the input parameters are $m_{o}$,
$m_{1/2}$, $A_{o}$, $\beta (M_{Z})$ and $m_{t} (M_{Z})$.

Since radiative corrections are generally expected to be small with the
exception of the contributions from the top-stop system, a reasonable
approximation of (7) is
\begin{equation}
\Delta V_{1} = {\frac {3} {32 {\pi}^{2}}} \sum_{i=+,-} m_{i} ^{4}
[ln(m_{i}
 ^{2}/Q^{2}) - 3/2]-{\frac {3} {16 {\pi}^{2}}} m_{t} ^{4}[ln(m_{t}
 ^{2}/Q^{2})-3/2]
\end{equation}
where we have kept only the $\tilde{t}$,$\tilde{t^{c}}$ and t
contributions$^{\cite{castano} \cite{zwirner} \cite{tracas}}$. The
field-dependent ``masses" appearing in (12) are
\begin{eqnarray}
  m_{t} \equiv Y_{t} H_{2} ^{o}\quad ,\quad
m_{\pm} ^{2} ={\frac {1} {2}}\{ m_{LL} ^{2} + m_{RR} ^{2} \pm [(m_{LL}
 ^{2}-m_{RR} ^{2}) ^{2} + 4 m_{RL} ^{4}] ^{\frac {1} {2}} \}
\end{eqnarray}
 where
\begin{eqnarray}
m_{LL} ^{2}& \equiv& m_{\tilde{t}}^{2} +m_{t} ^{2}+({\frac {1} {12}}
g'^{2}
-{\frac {1} {4}} g^{2})(|H_{2} ^{o} |^{2}-|H_{1} ^{o}|^{2})\nonumber\\
 m_{RR} ^{2} &\equiv& m_{\tilde{t^{c}}} ^{2} + m_{t} ^{2} -{\frac {1}
{3}}
{g'^{2}} ( |H_{2} ^{o}| ^{2} - |H_{1} ^{o}| ^{2})\nonumber\\
 m_{LR} ^{2} &\equiv& -Y_{t} (H_{2} ^{o} A_{t} + {\mu H_{1}^{o}}^{\ast})
 \\ m_{RL} ^{2} &\equiv& {m_{LR}^{2}}^{\ast}
\end{eqnarray}
Note that the use of the $t$-$\tilde{t}$ contribution
can be misleading in some cases due to large cancellations occuring
with terms that are not included [Arnowitt and Nath in ref.6].A complete
 analysis requires that the
 contributions of all sectors to the  effective potential are duly
taken into
account. \\
\\
\\
{\bf 4.  The Renormalization Group and Threshold effects}
\\

Consider the Renormalization Group equation for a soft mass parameter
derived
in the $\overline{DR}$ scheme$^{\cite{siegel}}$
\begin{equation}
{\frac {dm} {dln{Q^{2}}}}={\frac {b} {16 {\pi}^{2}}} m
\end{equation}
This equation should be integrated from a superlarge scale $Q=M_{GUT}$,
 where we
impose a boundary condition $m(M_{GUT})=m_{o}$, down to any desirable
value of
Q. As we come down from $M_{GUT}$ as long as we are at scales larger
than the
heaviest particle in the spectrum we include in $b$ contributions from
all the
particles in the MSSM. When we cross the heaviest particle threshold we
 switch
and compute $b$ in a new theory, an effective field
theory$^{\cite{Weinberg}}$ with the heaviest
particle integrated out. Coming further down in energy we encounter
the next
particle threshold at which point we switch again to a new effective
field
theory with the two heaviest particles integrated out. It is clear how
we
procced from then on.

The change in the running mass parameter $m$ at a particle
threshold $M$ in the
above scheme comes out to be for $m<M$
\begin{equation}
{\frac {\Delta{m}} {m}} \simeq {\frac {(b_{+} -b_{-})} {16 {\pi} ^{2}}}
ln({\frac {M^{2}} {{m}^{2}} })
\end{equation}
where $b_{+}$ and $b_{-}$ are the Renormalization Group coefficients
computed
in the effective theories above and below the threshold respectively.
Comparing
(17) with the exact result obtained from the 2-point function
associated
with $m$ we find that there is a finite non-logarithmic part that
is missed by
our approximation$^{\cite{Pierce}}$. The further $M$ and $m$ are apart
 the better the
approximation becomes.Of course, the great advantage of the
 approximation
lies in the fact that it is done entirely at the level of the
 Renormalization
Group without the need to calculate the finite parts of n-point
functions.

The Renormalization Group equation (16) referring to a particular
running
mass $m(Q)$ is integrated stepwise in the above stated manner down to
the
physical mass corresponding to $m(Q)$. The physical mass is determined
 by the
condition
\begin{equation}
m(m_{phys}) = m_{phys}
\end{equation}
Note that in the $\overline{DR}$ scheme the inverse two-point function
corresponding to the running mass $m(Q)$ will be at 1-loop of the
general form
\begin{equation}
{\Gamma}^{(2)} (Q^{2} , m(Q)) = Q^{2} (c_{1}+c_{2}
ln({\frac {m^{2} (Q)}
{Q^{2}}})) + m^{2} (Q) (c_{1}' + c_{2}'
ln({\frac {m^{2} (Q)} {Q^{2}}}))
\end{equation}
Imposing the condition (18) we see that the right hand side of (19)
 gives
$(c_{1}+c_{1}') m_{phys} ^{2}$.$\newline$ Thus, condition (18)
coincides with the
true (pole) condition for the physical mass only when the constant
non-logarithmic contributions can be neglected.

In what follows we shall present the 1-loop $\beta$-functions of
gauge and
Yukawa couplings as well as those for the soft masses, cubic parameters
$A$
and quadratic parameters $B$ and $\mu$$^{\cite{tamvakis}}$. Note that
the threshold corrections
introduced in our approximation by the theta-functions at 1-loop are
expected to be comparable to the
standard 2-loop RG corrections. In our numerical analysis that will
follow we shall employ the 2-loop RG
equations which have not been presented here due to their complicated
form but can be found
elsewhere$^{\cite{martin}}$.
In our notation, for a physical mass $M$,
\begin{equation} \theta_{M} \equiv \theta (Q^{2} -M^{2})
\end{equation}
Also $t$ stands for $t=ln Q^{2} $ and $\beta_{\lambda} \equiv {\frac
{d{\lambda}} {dt}}$ for each parameter $\lambda$. Note also that we
assume
diagonal couplings in family space.

The $\beta$-functions for the three gauge couplings are
\begin{eqnarray}
{\frac {dg_i}{dt}}&\equiv&{\beta(g_i)} =
{\frac {b_i}{ {2{(4\pi)}^2 }}  }
T_i\,{g_i}^3 \quad , \quad i=1,2,3
\end{eqnarray}
The coefficients $b_i$ are $\frac {33} {5}$,$1$,$-3$ respectively.

Keeping the Yukawa couplings $Y_{t,b,\tau}$ of the third generation
fermions, the corresponding $\beta$ functions are,
\begin{eqnarray}
{\frac {dY_\tau}{dt}}& \equiv& {\beta(Y_\tau)}={\frac {Y_\tau}{{(4\pi)}^2} }
  \{- {3 \over 2}T_{\tau2} {g_2}^2 - {\frac {9}{10}}T_{\tau1} {g_1}^2
  +2 T_{\tau\tau} {Y_\tau}^2 + {3 \over 2} {Y_b}^2  \}  \\
{\frac {dY_b}{dt}}& \equiv& {\beta(Y_b)}=  \nonumber \\
& &{\frac {Y_b}{{(4\pi)}^2} }
  \{-{8\over3}T_{b3} {g_3}^2 - {3 \over 2} T_{b2} {g_2}^2
  -{\frac {7}{30}}T_{b1} {g_1}^2
+{1 \over 2} T_{bt} {Y_t}^2 + 3T_{bb} {Y_b}^2+ {1 \over 2} {Y_\tau}^2 \}  \\
{\frac {dY_t}{dt}}& \equiv& {\beta(Y_t)}={\frac {Y_t}{{(4\pi)}^2} }
  \{-{8\over3}T_{t3} {g_3}^2-{3 \over 2}  T_{t2} {g_2}^2
  - {\frac {13}{30}}T_{t1} {g_1}^2
  + 3T_{tt} {Y_t}^2 + {1 \over 2}T_{tb} {Y_b}^2  \}
\end{eqnarray}
The threshold coefficients $T_i,\, T_{{\tau}i}, etc$ appearing in the
 expressions
above  are shown in Table I. We denote by $\tilde{G}$,$\tilde{W}$,
$\tilde{B}$ the
$SU(3)$,$SU(2)$ and $U(1)$ gauge fermions respectively.

The $\beta$-functions for the cubic couplings are
\begin{eqnarray}
 {\frac {dA_\tau}{dt} }&=&{\frac {1}{{(4\pi)}^2} }   \{
-3{g_2}^2 {M_2} \tt{W}{H_1} -{3 \over 5}{g_1}^2 {M_1}
{(2+\t{H_1})\t{B}}  \nonumber  \\  \nonumber  \\
&+&3{Y_b}^2 {A_b} \tt{D}{Q}+4{Y_\tau}^2 {A_\tau}+{A_\tau}[ {Z_{\tau 1}}
{g_1}^2
+ {Z_{\tau 2}}{g_2}^2 +{Z_{\tau \tau}}{Y_\tau}^2]    \}   \\  \nonumber
 \\
 {\frac {dA_b}{dt} }&=&{\frac {1}{{(4\pi)}^2} }   \{
-{16 \over 3}{g_3}^2 {M_3} \t{G} -3{g_2}^2 {M_2} \tt{W}{H_1}
-{1 \over 30}{g_1}^2 {M_1} {(-4+18 \t{H_1}) \t{B} }  \nonumber \\
\nonumber \\
&+&{Y_\tau}^2 {A_\tau} \tt{E}{L}
+ {Y_t}^2 {A_t} \st{H_2}{U}
+6{Y_b}^2 {A_b}        \nonumber  \\  \nonumber  \\
&+& {A_b}[{Z_{b3}} {g_3}^2+{Z_{b2}} {g_2}^2+{Z_{b1}} {g_1}^2
+ {Z_{bt}} {Y_t}^2   +   {Z_{bb}} {Y_b}^2  ]    \}     \\ \nonumber  \\
 {\frac {dA_t}{dt} }&=&{\frac {1}{{(4\pi)}^2} }   \{
-{16 \over 3}{g_3}^2 {M_3} \t{G} -3{g_2}^2 {M_2} \tt{W}{H_2}
-{1 \over 15}{g_1}^2 {M_1} {(4+9 \t{H_2}) \t{B} }  \nonumber \\
\nonumber \\
&+& 6{Y_t}^2 {A_t} \st{H_1}{U}
+{Y_b}^2 {A_b}\st{H_1}{D}         \nonumber  \\  \nonumber  \\
&+& {A_t}[{Z_{t3}} {g_3}^2+{Z_{t2}} {g_2}^2+{Z_{t1}} {g_1}^2
+ {Z_{tt}} {Y_t}^2   +   {Z_{tb}} {Y_b}^2  ]    \}     \\
\nonumber
\end{eqnarray}
In our notation $\theta_{ab} \equiv \theta_a \theta_b$ and
$\theta_{abc} \equiv
\theta_a \theta_b \theta_c$. The coefficients $Z_{qi}$ are shown in
Table II.

Next we procceed to the RG equations for the scalar masses. The
RG equations for
the sparticle masses refer to the third generation. For the other
two
generations the Yukawa couplings could be set to zero due to their
smallness.
\begin{eqnarray}
{\frac {d{\mm{Q}}} {dt}  }      
&=&  {\frac {1}{{(4\pi)}^2} }  \{- [{8\over3}{g_3}^2(\t{Q}-\t{G})
+{3\over2}{g_2}^2(\t{Q}-\t{W})
+{1 \over 30}{g_1}^2(\t{Q}-\t{B}) ] \, {\mm{Q}}
\nonumber \\  \nonumber  \\
&-&{16\over3}{g_3}^2 {M_3^2} \t{G}-3{g_2}^2 {M_2^2} \t{W}
-{1 \over 15}{g_1}^2 {M_1^2} \t{B} + {1 \over 10}{{g_1}^2} S
\nonumber \\ \nonumber \\
&+&{Y_t}^2 [\mm{Q} \t{H_2} +\mm{U} \t{U}+\smm{2} \s{H_2}+A_t^2
\st{H_2}{U}
	+ {\mu^2} (\st{H_1}{U}-2\t{H_2})]   \}   \nonumber \\ \nonumber
 \\
&+&{Y_b}^2 [\mm{Q} \t{H_1} +\mm{D} \t{D}+\smm{1} \s{H_1}+A_b^2
\st{H_1}{D}
	+ {\mu^2} (\st{H_2}{D}-2\t{H_1}) ]   \}  \\ \nonumber  \\
{\frac {d{\mm{U}}} {dt}  }      
&=&  {\frac {1}{{(4\pi)}^2} }  \{-[{8\over3}{g_3}^2(\t{U}-\t{G})
+{8 \over 15}{g_1}^2(\t{U}-\t{B}) ] \, {\mm{U}}
\nonumber \\  \nonumber  \\
&-&{16\over3}{g_3}^2 {M_3^2} \t{G}
-{16 \over 15}{g_1}^2 {M_1^2} \t{B} - {2 \over 5}{{g_1}^2} S
\nonumber \\ \nonumber \\
&+&2{Y_t}^2 [\mm{U} \t{H_2} +\mm{Q} \t{Q}+\smm{2} \s{H_2}+A_t^2
\st{H_2}{Q}
	+ {\mu^2} (\st{H_1}{Q}-2\t{H_2})]   \}   \\ \nonumber \\
{\frac {d{\mm{D}}} {dt}  }      
&=&  {\frac {1}{{(4\pi)}^2} }  \{- [{8\over3}{g_3}^2(\t{D}-\t{G})
+{2 \over 15}{g_1}^2(\t{D}-\t{B}) ] \, {\mm{D}}
\nonumber \\  \nonumber  \\
&-&{16\over3}{g_3}^2 {M_3^2} \t{G}
-{4 \over 15}{g_1}^2 {M_1^2} \t{B} + {1 \over 5}{{g_1}^2} S
\nonumber \\ \nonumber \\
&+&2{Y_b}^2 [\mm{D} \t{H_1} +\mm{Q} \t{Q}+\smm{1} \s{H_1}+A_b^2
\st{H_1}{Q}
	+ {\mu^2} (\st{H_2}{Q}-2\t{H_1})]   \}   \\ \nonumber \\
{\frac {d{\mm{L}}} {dt}  }      
&=&  {\frac {1}{{(4\pi)}^2} }  \{- [
{3\over2}{g_2}^2(\t{L}-\t{W})
+{3 \over 10}{g_1}^2(\t{L}-\t{B}) ] \, {\mm{L}}
\nonumber \\  \nonumber  \\
&-&3{g_2}^2 {M_2^2} \t{W}
-{3 \over 5}{g_1}^2 {M_1^2} \t{B} - {3 \over 10}{{g_1}^2} S
\nonumber \\ \nonumber \\
&+&{Y_\tau}^2 [\mm{L} \t{H_1} +\mm{E} \t{E}+\smm{1} \s{H_1}
+A_\tau^2  \st{H_1}{E}
	+ {\mu^2} (\st{H_2}{E}-2\t{H_1})]   \}   \\ \nonumber \\
{\frac {d{\mm{E}}} {dt}  }      
&=&  {\frac {1}{{(4\pi)}^2} }  \{- [
{6 \over 5}{g_1}^2(\t{E}-\t{B}) ] \, {\mm{E}}
\nonumber \\  \nonumber  \\
&-&{12 \over 5}{g_1}^2 {M_1^2} \t{B} + {3 \over 5}{{g_1}^2} S
\nonumber \\ \nonumber \\
&+&2{Y_\tau}^2 [\mm{E} \t{H_1} +\mm{L} \t{L}+\smm{1} \s{H_1}
+A_\tau^2  \st{H_1}{L}
	+ {\mu^2} (\st{H_2}{L}-2\t{H_1})]   \}   \\ \nonumber \\
{\frac {d{\smm{1}}} {dt}  }      
&=&  {\frac {1}{{(4\pi)}^2} }  \{- [
{3\over2}{g_2}^2(\s{H_1}-\tt{H_1}{W})
+{3 \over 10}{g_1}^2(\s{H_1}-\tt{H_1}{B} ) ] \, {\smm{1}}
\nonumber \\  \nonumber  \\
&-&3{g_2}^2( {M_2^2+\mu^2}) \tt{H_1}{W}
-{3 \over 5}{g_1}^2( {M_1^2+\mu^2}) \tt{H_1}{B}-{3 \over 10}{{g_1}^2}
 S
\nonumber \\ \nonumber \\
&+&{Y_\tau}^2 [\smm{1} +\mm{L} \t{L}+\mm{E} \t{E}+A_\tau^2
\tt{L}{E} ]
\nonumber \\ \nonumber \\
&+&3{Y_b}^2 [\smm{1} +\mm{Q} \t{Q}+\mm{D} \t{D}+A_b^2  \tt{Q}{D} ]
+ 3{{Y_t}^2} {\mu^2} \tt{Q}{U}
   \}   \\ \nonumber \\
{\frac {d{\smm{2}}} {dt}  }      
&=&  {\frac {1}{{(4\pi)}^2} }  \{- [
{3\over2}{g_2}^2(\s{H_2}-\tt{H_2}{W})
+{3 \over 10}{g_1}^2(\s{H_2}-\tt{H_2}{B} ) ] \, {\smm{2}}
\nonumber \\  \nonumber  \\
&-&3{g_2}^2( {M_2^2+\mu^2}) \tt{H_2}{W}
-{3 \over 5}{g_1}^2( {M_1^2+\mu^2}) \tt{H_2}{B}+{3 \over 10}{{g_1}^2}
 S
\nonumber \\ \nonumber \\
&+&3{Y_t}^2 [\smm{2} +\mm{Q} \t{Q}+\mm{U} \t{U}+A_t^2  \tt{Q}{U} ]
\nonumber \\ \nonumber \\
&+&3{Y_b}^2 {\mu^2} \tt{Q}{D}
+ {{Y_\tau}^2} {\mu^2} \tt{E}{L}       \}
 \\ \nonumber
\end{eqnarray}

The quantity $S$ appearing in the equations above is defined as
\begin{equation}
S \equiv \hs   Tr \hs  \{ {Y\over 2} \hs  \theta_m \hs  m^2 \}
\end{equation}
In the absence of the threshold effects this quantity is multiplicatively
renormalized. Therefore if it vanishes at the unification scale, due to
appropriate boundary conditions, it vanishes everywhere and its effect can be
ommited altogether from the RGE's. However in our case this does not any
longer hold owing to its explicit threshold dependence and $S$ starts
becoming  nonvanishing as soon as we pass the heaviest of the thresholds.
 For the Higgs and Higgsino mixing parameters $m_3^2 \equiv B \mu$
and $\mu$ respectively we {\mbox{have,}}
\begin{eqnarray}
{\frac {d{\smm{3}}} {dt}  }      
&=&  {\frac {1}{{(4\pi)}^2} }  \{ [
-{3 \over 4}{g_2}^2
{ (\s{H_1}+\s{H_2}+2\ss{H_1}{H_2}-\tt{H_1}{W}-\tt{H_2}{W})}
\nonumber \\  \nonumber  \\
&-&{3 \over 20}{g_1}^2
 { (\s{H_1}+\s{H_2}+2\ss{H_1}{H_2}-\tt{H_1}{B}-\tt{H_2}{B})}
\nonumber \\  \nonumber  \\
&+&{3 \over 2}{Y_t}^2+{3 \over 2}{Y_b}^2 +{1 \over 2}{Y_\tau}^2 ]
\,\,\, {\smm{3}}
\nonumber \\  \nonumber  \\
&+& \mu \, [- 3{g_2}^2 {M_2} \ttt{H_1}{H_2}{W}
-{3 \over 5}{g_1}^2{M_1} \ttt{H_1}{H_2}{B}
\nonumber \\  \nonumber  \\
&+&3 A_t{Y_t}^2 \tt{Q}{U}+3 A_b{Y_b}^2 \tt{Q}{D}
+A_\tau {Y_\tau}^2 \tt{L}{E} ]  \}
\end{eqnarray}
\begin{eqnarray}
{\frac {d{\mu}} {dt}  }      
&=&  {\frac {1}{{(4\pi)}^2} }  \{
{3 \over 8}{g_2}^2 (\t{H_1}+\t{H_2}-8\tt{H_1}{H_2}+\st{H_1}{W}+
\st{H_2}{W})
\nonumber \\  \nonumber  \\
&+&{3 \over 40}{g_1}^2(\t{H_1}+\t{H_2}-8\tt{H_1}{H_2}+\st{H_1}{B}+
\st{H_2}{B})
\nonumber \\  \nonumber  \\
&+&{3 \over 4}{Y_b}^2(\t{Q}+\t{D})+{3 \over 4}{Y_t}^2(\t{Q}+\t{U})
+{1 \over 4}{Y_\tau}^2(\t{L}+\t{E}) \} \, \mu
\end{eqnarray}
Finally the beta functions for the three gaugino masses are
\begin{eqnarray}
{\frac {d{M_i}} {dt}  }      
={S_i}\hs {\frac {b_i}{{(4\pi)}^2} } \hs {g_i^2} \hs { M_i} \hs ,
\hs i=1,2,3
\end{eqnarray}
where $b_i$ are the beta function coefficients of the gauge couplings
given
earlier
and $S_i$ are threshold function coefficients given by,
\begin{eqnarray}
S_3&=&-3 \hs  \t{G}
-{1 \over 6} \hs{\sum_{i=1}^{N_g}} \hs (2\t{Q_i}+\t{U_i}+\t{D_i})
\\ \nonumber \\
S_2&=& -6 \hs \t{W}-{1 \over 2} \hs {\sum_{i=1}^{N_g}} \hs (3\t{Q_i}+
\t{L_i})
-{1 \over 2}(\st{H_1}{H_1}+\st{H_2}{H_2})
\\ \nonumber \\
S_1&=&{1 \over 11}\hs [\hs {\sum_{i=1}^{N_g}}  \hs
({1 \over 6}\t{Q_i}+{4 \over 3}\t{U_i} +
{1 \over 3}\t{D_i}+{1 \over 2}\t{L_i}+\t{E_i})
+{1 \over 2}(\st{H_1}{H_1}+\st{H_2}{H_2})  ]
\end{eqnarray}
The dimensionful parameters, masses and cubic couplings, are meant to
freeze out when the energy crosses below the mass scale associated with the
heaviest particle participating. This can be implemented by multiplying the
corresponding quantity by the relevant theta function. Thus for instance
$A_t$ freezes out below the thresholds of either $\tilde{t}$,
or $\tilde{t} ^{c}$, or $H_2$, whichever is the heaviest,
and the associated  theta functions should multiply
the r.h.s of eq. (27). For simplicity of our notation we do not indicate
that explicitly in the RGE's displayed in eqs. (25)-(38).
\\
\\
\\
{\bf 5.  Formulation of the problem and numerical analysis}
\\

The problem at hand consists in finding the physical masses of the
presently
unobserved particles, i.e. squarks, sleptons, Higgses, Higgsinos and
gauginos, as well as their physical couplings to other observed
particles. This
will be achieved by integrating the Renormalization Group equations
from a
superheavy scale $M_{GUT}$, taken to be in the neighbourhood of $10^{16}
GeV$, down
to a scale $Q_o$ in the stepwise manner stated. If the equation at
 hand is the
Renormalization Group equation for a particular running mass $m(Q)$,
then $Q_o$
is the corresponding physical mass determined by the condition
$m(Q_o)=Q_o$. If the equation at hand is the Renormalization Group
equation for a gauge or Yukawa coupling the
integration will be continued down to $Q_o=M_Z$. Acceptable solutions
should
satisfy the constraints (8) and (9) at $M_Z$, i.e. describe a
low energy
theory with broken electroweak symmetry at the right value of
 $M_Z \simeq
91.187 GeV$.

The boundary condition at high energy will be chosen as simple as
possible, postponing for elsewhere the study of more complicated
alternatives. Thus at the (unification) point $M_{GUT}$, taken to be
 $10^{16}$ GeV, we shall take
\begin{eqnarray}
m_{\tilde{Q}} (M_{GUT})& =& m_{\tilde{D^c}}
(M_{GUT}) = m_{\tilde{U}^c} (M_{GUT}) =
m_{\tilde{L}} (M_{GUT}) = m_{\tilde{E}^c} (M_{GUT})\nonumber
\\ & =& m_{H_1} (M_{GUT}) =
m_{H_2} (M_{GUT}) \equiv m_o
\end{eqnarray}
and
\begin{equation}
M_1 (M_{GUT}) = M_2 (M_{GUT}) = M_3 (M_{GUT}) \equiv m_{1/2}
\end{equation}
In addition we take equal cubic couplings at $M_{GUT}$, i.e.
\begin{equation}
A_e (M_{GUT}) = A_d (M_{GUT}) = A_u (M_{GUT}) \equiv A_o
\end{equation}
All our boundary conditions are family blind. We shall also denote
 with $B_o$
and $\mu_o$ the boundary values at $M_{GUT}$ of the parameters $B(Q)$
 and
$\mu(Q)$. This five parameters $m_o$, $m_{1/2}$, $A_o$, $B_o$ and
 $\mu_o$ are not
all free due to conditions (8) and (9) which could be viewed as
determining
$B$ and as trading $\mu$ for $\beta \equiv tan^{-1}
(\upsilon_2/\upsilon_1)$. Thus, $m_o$, $m_{1/2}$, $A_o$, $\beta(M_Z)$
as well as the sign of $\mu(M_Z)$ can be
our free parameters.

Our set of constraints includes the low energy experimental gauge
coupling values which we have taken
to be
${\alpha_3 (M_Z)}_{\overline {MS}}=0.117 \pm 0.010$,
${{\alpha_{em}}(M_Z)^{-1}}_{\overline
{MS}}=127.9\pm 0.1$ and  $(sin^{2}{\theta_W})\mid_{\overline
 {MS}}=0.2316-.88\;10^{-7} ({M_t}^2-160^2) GeV^{-2}$$^{\cite{polonsky}}$.
 The $\overline {MS}$ values for the couplings
are related to their  $\overline {DR}$\footnote[1] {Note that at the
 2-loop order the $\overline {DR}$
scheme needs to be modified so that no contribution to the scalar
 masses due to
the``$\epsilon$-scalars"$^{\cite{jack}}$ shows up.}ones through the
relations
$g_{\overline {MS}}=g_{\overline{DR}}(1- C g^2/{96 \pi^{2}})$,
where $C=0,2,3$ respectively for the three gauge groups. The
unification scale $M_{GUT}$ is determined from the intersection of
${\alpha_1}_{\overline{DR}}$ and ${\alpha_2}_{\overline{DR}}$
 gauge couplings
and is found to be in the vicinity of $10^{16}$ GeV.This value of $M_{GUT}$
is not easily reconcilable with the low energy value of $\alpha_3$ quoted above
and the universal boundary condition, as given in eqs.(42-44) if the effective
SUSY breaking scale $M_S$ is  below 1 Tev$^{\cite{rozkowsi}}$
If we treat the low energy value of $\alpha_3$ as an output we find it
to be $\geq$0.125, i.e slightly larger than the most favourite experimental
value quoted previously with a tendency to decrease as $M_S$ gets larger than
$\geq$1 Tev.Our interest in this paper is mainly focused on the mass spectrum
and on the effect of the mass  thresholds to it.The subtle issue of the gauge
coupling unification in  conjuction      with the small value of $\alpha_3$
shall be addressed to in a forthcoming publication. In the course of our
numerical computations we allow for switching off the effect of the the
thresholds from masses and cubic couplings involved. In this way we can
compare the predictions for couplings and pole masses in the two cases
i) With all thresholds present in both couplings and dimensionful parameters
and ii) thresholds appearing only in gauge and Yukawa couplings. The latter
case has already been considered by several groups. Although the difference
 is expected to be small only case  (i) represents a consistent prediction of
the spectrum, together with $\alpha_{GUT}$ and $M_{GUT}$, in the framework of
the leading logarithmic approximation.

For the $b$ quark and  $\tau$ lepton our inputs are $M_b^{pole}=5$ GeV and
$M_{\tau}^{pole}=1.777$ GeV which are related to their running
$\overline {MS}$ masses by
\begin{equation}
m_b(M_b)=\frac {M_b} {(1+\frac {4\alpha_3} {3 \pi} +12.4 ({\frac
 {\alpha_3} {\pi})^{2}})}
\end{equation}
\begin{equation}
m_{\tau} (M_{\tau})=M_\tau
\end{equation}
These are evolved up to $M_Z$ according to the $SU_{c} (3)\times U_{em} (1)$
 RGE's and are then converted to $\overline {DR}$ values.From these the
$\overline {DR}$ values for the Yukawa couplings $Y_b (M_Z)$, $Y_\tau (M_Z)$
are determined.
The numerical procedure for the determination of $M_{GUT}$
respecting the experimental inputs for $\alpha_{em}$,$sin^{2}\theta_{W}$
and $M_{b,\tau} ^{pole}$ needs several iterations to reach convergence.

As for the top Yukawa coupling our input is the $\overline {DR}$ running
mass $m_{t} (M_Z)$ related to the physical top mass $M_t ^{pole}$ by
\begin{equation}
m_t (M_t)= \frac {M_t} {(1+ \frac {5 \alpha_3} {3 \pi} +\cdots)}
\end{equation}
 The recent evidence$^{\cite{abe}}$ for the
top-quark mass has motivated values for $M_t$ in
the neighborhood of $176 \pm 8$ GeV.

In our numerical procedure we follow a two loop renormalization
group analysis for all parameters involved, i.e couplings and dimensionful
parameters, v.e.v's included. We start with the $\overline {MS}$ values for
the gauge couplings at $M_Z$ giving a trial input value for the strong
coupling constant ${\alpha_3}$ in the vicinity of $.120$, which are then
converted to their $\overline {DR}$ values. These are run down to
$M_b , M_{\tau}$ with the $SU_{c} (3)\times U_{em} (1)$  RGE' s to know the
running bottom and tau masses as given before. From these by use of the
RGE's for the running masses  $m_b, m_{\tau}$
we run upwards to know their $\overline {MS}$
values at $M_Z$ which are subsequently converted to their corresponding
$\overline {DR}$ values. This provides us with the bottom and tau Yukawa
couplings at the scale $M_Z$. The top Yukawa coupling is known from the
input running top quark mass $m_t(M_Z)$ as said earlier.
The evolution of all couplings from $M_Z$ running upwards
to high energies determines the unification scale $M_{GUT}$
and the value of the unification coupling $\alpha_{GUT}$ by
\begin{equation}
{\alpha_1}_{\overline{DR}}(M_{GUT}) = {\alpha_2}_{\overline{DR}}(M_{GUT})
= {\alpha_{GUT}}.
\end{equation}
Running down from $M_{GUT}$ to $M_Z$ the trial input value for ${\alpha_3}$
has now changed. This procedure is iterated several times until convergence
is reached. In each iteration the values of $B,\mu$, which
as stated previously are not inputs in this approach, are
determined by minimizing the scalar
potential. For their determination at the scale $M_Z$ we take into account
the one loop corrected scalar potential.
This procedure modifies the tree level values $B(M_Z)$, $\mu(M_Z)$.
It is well known that the value of $\mu$ affects the predictions for the
physical masses especially those of the neutralinos and charginos. In
approaches in which the effect of the thresholds is ignored in the RGE's the
determination of $B,\mu$ is greatly facilitated by the near decoupling of
these parameters from the rest of the RGE's. However with the effects of the
thresholds taken into account such a decoupling no longer holds since the
thresholds themselves depend on $B,\mu$, or equivalently on $\mu,m_3^2$.

In solving the system of the 33 RGE's involved, among these those for the
v.e.v's, we have used special $FORTRAN$ routines of the $IMSL$ library
available to us which use the $Runge - Kutta - Verner$ sixth order method
and are capable of handling stiff systems of
nonlinear differential equations with high accuracy.

Throughout our analysis we avoid considering values for $\tan \beta$ for
which the couplings are driven to large values  outside  of the
perturbative regime.

The experimental lower bounds for the masses of new particles
extracted from accelerator data are as
follows\footnote[2] {We have also constrained the output to cases of
 neutral and colourless LSP.}.
The four neutralino mass eigenstates have to be heavier than
$20$,$45$,$70$ and $108$ GeV while the
two chargino states have lower mass limits of $43$ and $99$ GeV.
Charged sleptons have to be heavier
than $45$ GeV, while sneutrinos have to be heavier than 41.8 GeV. There
is also a $150$ GeV lower bound on the mass of squarks and gluinos. Charged
Higgses should be heavier than $41$ GeV while the CP-odd neutral Higgs
should be heavier than $22$ GeV. The lightest of the two neutral CP-even
Higgs eigenstates should have a mass larger than $44$  if the CP-odd
Higgs is lighter than $M_Z$ or $60$ GeV if the opposite is true
respectively.

The interesting part of the output of the numerical integration of the
 RG equations consists of the
mass spectrum of the new particles (gaugino-Higgsinos, squarks and
sleptons) as well as the Higgs
masses. The neutralino mass eigenstates can be read off from
reference ${\cite{tracas}}$.
The two stop mass eigenstates
$\tilde {t_1}$ and $\tilde {t_{2}}$ correspond to the mass
eigenvalues
$m_{\pm}^{2}$ shown in(13).The Higgs mass eigenvalues at tree
level are
$m_{A}^{2}=m_{1}^{2}+m_{2}^{2}$ for the neutral pseudoscalar Higgs A,
and
\begin{equation}
m_{H,h}^{2}={\frac {1} {2}}[{M_{Z}}^{2}+m_{A}^{2} \pm
\sqrt{(m_{A}^{2}+M_{Z}^{2})^{2}-4 M_{Z}^{2}
m_{A}^{2} cos^{2}(2 \beta)}] \nonumber\\
\end{equation}
for the two neutral scalar Higgses $h$ and $H$. The charged Higgs has a
 tree level mass
$M_{H^{+}}^{2}=m_{A}^{2}+M_{W}^{2}$. As is well known, the 1-loop
radiative corrections, mostly due to
the large value of the top Yukawa coupling, are important for the Higgs
 masses. Following an
approximation that has been tested in the literature$^{\cite{zwirner}}$,
 we have computed the Higgs
mass eigenvalues based on the 1-loop effective potential (12), where the
dominant third generation contribution has been kept.

The radiative corrections to the pseudoscalar and charged Higgses
are known to
have an explicit dependence on the scale $Q$
which cancels against the implicit $Q$ dependence of the mixing mass
${m_3^2}(Q)$ and $\sin{2{\beta (Q)}}$.
Following other authors$^{\cite {zwirner}}$ we choose as appropriate
scale for the evaluation of
their masses a scale $Q_0$ for which the effect of the radiative
corrections
is vanishingly small. Then $Q_0$ turns out to be in the vicinity of  the
heavier of the stops and in this regime the stops have not been
decoupled yet.
It is therefore permissible to consider their loop effects to the
effective potential as given in the references cited above.
Keeping only the dominant third generation corrections to the Higgs masses
the aforementioned cancellation among the $Q$ dependent pieces is rather
incomplete since the evolution of ${m_3^2}(Q)$ depends on the gaugino
masses as well (see $\it {eq.} (36)$). Therefore there is a residual scale
dependence, induced by the gauginos, whose effect on ${m_3^2}(Q)$ is not
guaranteed to be small especially for large values of the soft
mass $m_{1/2}$.

In our approach we have verified that the radiatively corrected Higgs masses
for the heavy Higgses
as these are calculated at the scale $M_Z$ and the corresponding masses at
$Q_0$ are very close to each other even for large values of the soft
mass $m_{1/2}$. This is due to the appearance of the thresholds within
the RGE of ${m_3^2}(Q)$ which properly takes care of the gaugino decoupling.
This would not have been the case in schemes
in which such a decoupling is not present. In those cases it is
required
that either the physical Higgs masses are evaluated as poles of
the one loop
propagators or the effect of the gaugino fields is duly taken into account
in the effective potential approach, for the effect of the radiative
corrections to be numerically insensitive to the choice of the scale.
This
subtle issue is under investigation and the results of this analysis
will appear in a forthcoming publication$^{\cite {dedes}}$.
\\
\\

{\bf 6.  Conclusions}
\\

We have displayed some of our results in tables III to VIII.
In all tables we have taken $\mu(M_Z)$  positive. The negative
$\mu(M_Z)$ values lead to qualitatively similar results. In the
first four tables (III - VI) for a fixed value of
$m_{t}(M_{Z})=175$ GeV,
we  present acceptable spectra that
have been obtained for different values of the four parameter $m_o$,
 $A_o$, $m_{1/2}$ and
$tan\beta(M_Z)$. In the table III, for a characteristic set of values
$A_o=400$ GeV, $m_o=300$
GeV and $m_{1/2}=200$ GeV we have varied $tan\beta$ between $2$ and
$25$. For values of $tan\beta$ larger than about 60
no electroweak breaking occurs in this case. Note the well
 known$^{\cite{nanopoulos}}$
approximate equality between the masses of one of the neutralinos and
one of the charginos. The lightest Higgs turns out to be heavier than the
$Z$ - boson and increases with increasing the value of $tan\beta$.
Althought not displayed, for negative $\mu$ its mass drops
below $M_Z$ for small values of the angle ${tan\beta \simeq 2}$.
In table IV, for fixed characteristic
values of $tan\beta=10$, $A_o=400$ GeV and $m_o=300$GeV,we vary
 $m_{1/2}$ between $75$ GeV and 700
GeV. The sensitivity of the whole spectrum on $m_{1/2}$ is apparent.
In table V, for fixed
$tan\beta=10, m_o=300$ GeV and $m_{1/2}=200$ GeV, we vary $A_o$ between
 $0$ and $800$ GeV. In this
case, as in all other acceptable cases, the LSP is a neutralino with a
 mass roughly independent of
$A_o$.
We have also included in this table negative values of $A_o$
in the range -800 GeV to
0. Note that in both cases the gluino mass is roughly stable.
In table VI we keep $tan\beta=10$, $A_o=400$GeV and $m_{1/2}=200$GeV fixed
and vary $m_o$ between 100 and 800 GeV.
In all cases shown the LSP is a neutralino.
In table VII we have presented spectra obtained, for fixed $m_o$,
$m_{1/2}$, $A_o$ and
$tan\beta$ values, when we vary the top quark mass.
In all cases displayed in the tables III - VIII
we have the approximate equality $m_{\chi_{2}^{o}}\simeq
m_{\chi_{2}^{c}}$$^{\cite{nanopoulos}}$ due to the fact that
$\mu(M_Z)$ turns out to be significantly heavier than $M_Z$.

As is apparent from the results displayed in tables IV to VI the value of
the strong coupling constant has the tendency to decrease with increasing
the supersymmetry breaking scale. In the table IV for instance keeping fixed
the values of $m_0, A_0$, $\alpha_3 (M_Z)$ takes values between
$\simeq .131$ and $\simeq .127$ if $m_{1/2}$ varies from $75$ to $700 GeV$.
The situation is similar for the other two cases of tables V and VI
where two of the soft SUSY breaking parameters are kept fixed and the third
varies from low to large values. One notices however that
the decrease in $\alpha_3 (M_Z)$ in these cases
is rather slower as compared to that
of table IV. Thus $\alpha_3 (M_Z)$ decreases faster in the direction of
increasing the soft gaugino mass $m_{1/2}$.

Finally in table VIII for a characteristic choice of the parameter values
we compare outputs for three dinstict cases. One loop predictions
(case [a]), two loop predictions with thresholds only in couplings
(case [b]), and the complete two loop case where thresholds appear in  the
RGE's of both couplings and dimensionful quantities (case [c]). Comparing
cases [a] and [b]  we see that as expected the value of the strong coupling
is increased  from its one loop value from $\simeq .118$ to $.133$. The
bulk of this increase, as is well known, is due to the two loop effects.
At the same time the value of the unification scale is increased too from
$2.1881 \quad 10^{16}$ to  $2.8876 \quad 10^{16}GeV$.
 Two loop effects
delay the merging of the $\alpha_1$ and $\alpha_2$ gauge couplings
resulting in larger values of $\alpha_3$. As can be seen from the
comparison of the two outputs, two loops and
threshold effects in couplings have a small effect on the mass spectrum.
Differences are small of the order of $2 \%$ or smaller in all sectors
except the neutralino and chargino states
${\tilde{\chi}}_{1,2}^{o}$, ${\tilde{\chi}}_{2}^{c}$,
whose masses are exactly
$M_1, M_2$ in the absence of electroweak symmetry breaking effects. In
these states we have differences of the order of $10 \%$ or so
originating mainly from the different evolutions of the soft gaugino masses
$M_{1,2}$   and the $\alpha_{1,2}$
gauge couplings in the 1 - loop and 2 - loop cases.

Switching on the the mass and cubic coupling thresholds produces a minor
effect, as can be seen by comparison of cases [b] and [c], except in the case
of the   heavy Higgses and those neutralino and chargino states that are nearly
Higgsinos, i.e. of mass $\simeq \mu$, for large values of the parameter $\mu$.
These states are labelled as
${\tilde{\chi}}_{3,4}^{o}$, ${\tilde{\chi}}_{1}^{c}$
In those cases relatively
large differences are observed. For the neutralinos and charginos
the large differences are attributed to the
value of the parameter $\mu (M_Z)$ which turns out to be larger, by about
$10 \%$, in case [c] as compared to that of case [b]. As far as the heavy
Higgses are concerned from the discussion in the previous section
it is evident that this discrepancy is due
mainly to the  evolution of $m_3^2$, whose value affects substantially
the masses of the pseudoscalar and charged Higgses,
and in  particular on its dependence on the gaugino masses. In order to
determine the  Higgs masses we need, among other things, the value of $m_3^2$
at a scale $Q_o$, which minimizes the loop corrections due to the quarks and
squarks of  the third family, knowing its value at $M_Z$. However its
evolution in  the two approaches, with and without thresholds in masses
and cubic couplings, is quite
different. In the complete 2 - loop case, which takes into account all
threshold
effects, we have  properly  taken care of the decoupling of all particles as
we move below their  thresholds, gauginos included, unlike the case where the
threshold effects  are totally ignored. This results in rather large
logarithmic corrections  and hence to the relatively large differences
($\simeq 6 \%$) occuring in this cases. For a proper treatment of the
radiative  corrections in the second scheme, where threshold effects are not
present  in the RGE's, explicit one loop calculations of the gaugino
contributions to the two point Green's functions of the Higgses should be
carried out.  It is only in this case where  comparison of the predictions
for the  heavy Higgses spectrum of the two approaches at hand can be made. At
any rate the rather large differences seen in some particular cases
point to the fact that a more
refined analysis of the radiative effects to the Higgs sector is needed which
also takes into account of the contributions of the gauginos and not just those
of the heavy quarks. We have undertaken such a calculation and the results
will  appear in a future publication$^{\cite {dedes}}$.
\\
\\
\newpage
{\bf Acknowledgements} \\
We thank the CERN Theory Division for hospitality during a short
visit
in which part of this work was completed. K.T. also acknowledges
illuminating conversations with C. Savoy and I. Antoniadis during a
visit
at Saclay in the framework of the EEC Human Capital and Mobility
Network
"Flavourdynamics" (CHRX-CT93-0132). A.B.L.  acknoweledges support by
the
EEC Science Program SC1-CT92-0792.
Finally we all thank the Ministry of Research and Technology for
partial funding
of travelling to CERN.

\newpage


\baselineskip=20pt
\newpage
\noindent
{\bf Table Captions}

\vspace{1.cm}
\noindent
{\bf Table I}:\quad Threshold coefficients appearing in the renormalization
group equations \\
of the gauge and Yukawa couplings. Above all thresholds these become
equal to unity.

\vspace{1.cm}
\noindent
{\bf Table II}:\quad Threshold coefficients appearing in the renormalization
group equations \\
of the trilinear scalar couplings. Above all thresholds these are
vanishing.

\vspace{1.cm}
\noindent
{\bf Table III}:\quad MSSM predictions for $m_t=175 \,GeV,\,
A_o=400 \, GeV$,$\,m_o=300$,
\,${m_{1/2}}=200 GeV$ and for values of $\tan \beta$ ranging from 2 to 25.
Only the $\mu > 0$ case is displayed. The $ {\overline {MS}}$ values of
$\alpha_{em}$,
$sin^{2}{\theta_W}$, $ \alpha_{3}$ and the unification scale are
also shown.
$\alpha_{GUT}$ is the $ {\overline {DR}}$ value of the unification coupling.
$M_t$ is the physical top quark mass, $\tilde{g}$ denotes the
gluino and ${\tilde{\chi}}_{i}^{o}$, ${\tilde{\chi}}_{i}^{c}$ are
neutralino and chargino states. $({\tilde{t}}_{1,2},{\tilde{b}}_{1,2})$ and
$({\tilde{\tau}}_{1,2}, {\tilde{\nu}}_{\tau})$
are squarks and sleptons of the
third generation while  ${\tilde{u}}_{i}$,${\tilde{u}}_{i}^{c}$,
${\tilde{d}}_{i}$,${\tilde{d}}_{i}^{c}$ and
${\tilde{e}}_{i}$,${\tilde{e}}_{i}^{c}$, ${\tilde{\nu}}_{i} $ denote
squarks and sleptons of the first two generations.
$h_o$,$H_o$, $A$, $H^{\pm}$ are the $CP$ even, $CP$ odd and the charged
Higgses respectively.

\vspace{1.cm}
\noindent
{\bf Table IV}:\quad Same as in table III
for $m_t=175 \,GeV,\,
A_o=400 \, GeV$,\,$\,m_o=300$,\,$\tan \beta =10$
and for values of $m_{1/2}$ ranging from $75 GeV$ to $700 GeV$ $(\mu > 0)$.

\vspace{1.cm}
\noindent
{\bf Table V}:\quad Same as in table III for $m_t=175 \,GeV$,\,
$m_o=300\,GeV$, \,
\,${m_{1/2}}=200 GeV$ and for values of $A_o$ \, $0,\,{\pm 300},\,{\pm 800}$
\,$GeV$ $(\mu>0)$.

\vspace{1.cm}
\noindent
{\bf Table VI}:\quad Same as in table III for $m_t=175\,GeV$,\,
$A_o=400\,GeV$,\,$m_{1/2}=200 \,GeV$,\,$\tan \beta=10$ and values of
$m_o$ from $100 GeV$ to $800 GeV$ $(\mu>0)$.

\newpage
\noindent
{\bf Table VII}:\quad Mass spectrum of the MSSM for input values
$m_o=400\,GeV$,\,${m_{1/2}}=300 GeV$,\,$A_o=200\,GeV$,\,$\tan \beta = 20$ and
values for the running top quark mass equal to $175,180,185,190$ and
$195 \, GeV$\,$(\mu>0)$ respectively.Couplings and masses shown
are as in table III.

\vspace{1.cm}
\noindent
{\bf Table VIII}:\quad MSSM mass spectrum for the inputs shown in the first
row $(\mu>0)$. We compare 1 - loop (case [a]), 2 - loop with thresholds in
couplings (case [b]) and complete  2 - loop predictions (case [c]) with
thresholds in both couplings and dimensionful parameters.

\newpage
\begin{center}
\begin{tabular}{l}\hline
\multicolumn{1}{c}{\bf TABLE I} \\ \hline \\
$T_1={1 \over 33} [ 20 +
 {{\theta_{{\tilde H}_1}}+{\theta_{{\tilde H}_2}} } +
 {1 \over 2} (  {{\theta_{{H}_1}}+{\theta_{{ H}_2}} } )+
 {\sum_{i=1}^{3}}( {1 \over 2}
 {\theta_{{\tilde L}_i}}+ {\theta_{{\tilde E}_i}}+
 {1 \over 6}{\theta_{{\tilde Q}_i}}+{4 \over 3}{\theta_{{\tilde U}_i}}+
{1 \over 3}{\theta_{{\tilde D}_i}}          ) ]  $
\\  \\
$T_2=-{10 \over 3}+{4 \over 3}{\theta_{{\tilde W}}} +
{ 1\over 3}({{\theta_{{\tilde H}_1}}+{\theta_{{\tilde H}_2}} }        )+
{1 \over 6} (  {{\theta_{{H}_1}}+{\theta_{{ H}_2}} } ) +
 {1 \over 6}{\sum_{i=1}^{3}}(
3{\theta_{{\tilde Q}_i}}+{\theta_{{\tilde L}_i}}) $ \\   \\
$T_3={7 \over 3}-{2 \over 3}{\theta_{{\tilde G}}}-{1 \over 18}
{\sum_{i=1}^{3}}(2{\theta_{{\tilde Q}_i}}+{\theta_{{\tilde D}_i}}+
{\theta_{{\tilde U}_i}})$\\  \\  \hline \\

${T_{\tau 2}}={1 \over 4}[ -1+4{\theta_{{ H}_1}}-2{\theta_{ {{\tilde H}_1}
{\tilde W}}}-{\theta_{  {\tilde L}{\tilde W} } } +
4 {\theta_{   {{\tilde H}_1} {\tilde L}{\tilde W}  }   }  ] $ \\  \\

${T_{\tau 1}}={1 \over 12}[ 11-4{\theta_{ {\tilde B}{\tilde E} }}
+8 {\theta_{ {\tilde B} {\tilde E}{{\tilde H}_1} } }
-2 {\theta_{   {\tilde B}{{\tilde H}_1} } }+4{\theta_{{ H}_1}}
-{\theta_{  {\tilde B} {\tilde L}   }}
-4{\theta_{  {\tilde B} {\tilde L}{{\tilde H}_1}  }  } ] $ \\ \\

${T_{\tau \tau}}={1 \over 8}[ 2+\tt{H_1}{E}
+3{\theta_{{ H}_1}}+2{\theta_{  {\tilde L}{{\tilde H}_1}  }   }
   ] $ \\  \\  \hline  \\

${T_{b3}}={1 \over 4}[6-{\theta_{{\tilde G}{\tilde D}}}
-{\theta_{   {\tilde G}{\tilde Q} } }]$ \\ \\

${T_{b 2}}={1 \over 4}[ -1+4{\theta_{{ H}_1}}-2{\theta_{   {{\tilde H}_1}
{\tilde W} }}  -{\theta_{  { \tilde Q} {\tilde W}  } } +
4 {\theta_{  {{\tilde H}_1}{\tilde Q}{\tilde W}  } }] $ \\  \\

${T_{b 1}}={1 \over 28}[-21-4{\theta_{   {\tilde B}{\tilde D}   }}
-18{\theta_{   {\tilde B}{{\tilde H}_1}  }  }
+24{\theta_{ {{\tilde H}_1}{\tilde D}{\tilde B}     }}
+36{\theta_{{H}_1}}-{\theta_{   {\tilde B}{\tilde Q}   }}
+12{\theta_{  {{\tilde H}_1}{\tilde Q}{\tilde B}  } }]$ \\ \\

${T_{bt}}={1 \over 2}[\s{H_2}+\tt{ H_2}{ U}] $  \\ \\

${T_{bb}}={1 \over 12}[6+\tt{D}{H_1}+3\s{H_1}+2\tt{Q}{H_1}] $ \\ \\ \hline \\

${T_{t3}}={1 \over 4}[6-\tt{G}{Q}-\tt{G}{U}]$ \\ \\

${T_{t2}}={1 \over 4}[-1+4\s{H_2}-2\tt{H_2}{W}-\tt{Q}{W}+4\ttt{H_2}{Q}{W}]$
\\  \\
${T_{t1}}={1 \over 52}[15-18\tt{H_2}{B}+36\s{H_2}-\tt{B}{Q}-12\ttt{B}{Q}{H_2}
-16\tt{B}{U}+48\ttt{B}{U}{H_2}] $ \\ \\

${T_{tt}}={1 \over 12}[6+3\s{H_2}+2\tt{H_2}{Q}+\tt{H_2}{U}] $ \\ \\

${T_{tb}}={1 \over 2}[\s{H_1}+\tt{H_1}{D} ]  $ \\  \\ \hline
\end{tabular}
\end{center}
\newpage
\begin{center}
\begin{tabular}{l}\hline
\multicolumn{1}{c}{\bf TABLE II} \\ \hline \\
${Z_{\tau 1}}={3 \over 40} [ 11 +
10\t{B}-8\t{E}-4\tt{B}{E}+8\ttt{B}{E}{H_1}+ $  \\ \\
\hspace{5cm}  $  2\s{H_1}
 -8\st{H_1}{E}-2\t{L}-\tt{B}{L} -8\tt{E}{L}-4\ttt{B}{H_1}{L}
+4\st{H_1}{L}  ]  $
\\  \\
${Z_{\tau 2}}={1 \over 8} [ -3 +
6\s{H_1}-6\t{L}-12\st{H_1}{L}+6\t{W}-3\tt{L}{W}+12\ttt{W}{H_1}{L}] $
\\ \\
${Z_{\tau \tau}}={1 \over 4} [ -16 +
+6\t{H_1}-\tt{H_1}{E}-3\s{H_1} +4\st{H_1}{E} +4\tt{E}{L}
-2\tt{H_1}{L}+ 8\st{H_1}{L}  ]  $ \\ \\ \hline \\
${Z_{b3}}={2 \over 3} [6 -
2\t{D}+4\t{G}-\tt{D}{G}-2\t{Q}-4\tt{D}{Q}-\tt{G}{Q} $ \\ \\
${Z_{b2}}={1 \over 8} [-3
+6\s{H_1}-6\t{Q}-12\st{H_1}{Q}+6\t{W}-3\tt{Q}{W}+12\ttt{W}{H_1}{Q}] $
\\ \\
${Z_{b1}}={1 \over 120} [-21
+10\t{B}-8\t{D}-4\tt{B}{D}+24\ttt{B}{D}{H_1}+18\s{H_1}-24\st{H_1}{D} $ \\ \\
\hspace{5cm} $-2\t{Q}-\tt{Q}{B}+8\tt{Q}{D}+12\ttt{B}{Q}{H_1}-12\st{H_1}{Q}] $
\\ \\
${Z_{bb}}={1 \over 4} [-24
+6\t{H_1}-\tt{H_1}{D}-3\s{H_1}+4\st{H_1}{D}+12\tt{Q}{D}-2\tt{Q}{H_1}
+8\st{H_1}{Q}]$
\\ \\
${Z_{bt}}={1 \over 4} [2\t{H_2}-\s{H_2}-\tt{U}{H_2} ] $
\\ \\ \hline   \\
${Z_{t3}}={2 \over 3} [6 -
2\t{Q}+4\t{G}-\tt{Q}{G}-2\t{U}-4\tt{U}{Q}-\tt{G}{U} ] $ \\ \\
${Z_{t2}}={1 \over 8} [-3
+6\s{H_2}-6\t{Q}-12\st{H_2}{Q}+6\t{W}-3\tt{Q}{W}+12\ttt{W}{H_2}{Q}] $
\\ \\
${Z_{t1}}={1 \over 120} [15
+34\t{B}+18\s{H_2}-2\t{Q}-\tt{B}{Q}-12\ttt{B}{Q}{H_2}+12\st{H_2}{Q}  $ \\ \\
\hspace{5cm}$-32\t{U}-16\tt{B}{U}+48\ttt{B}{U}{H_2}-48\st{H_2}{U}
-16\tt{U}{Q}  ]  $
\\ \\
${Z_{tb}}={1 \over 4} [2\t{H_1}-\s{H_1}-\tt{D}{H_1} ] $
\\ \\
${Z_{tt}}={1 \over 4} [-24 \st{H_1}{U}
+6\t{H_2}-\tt{H_2}{U}-3\s{H_2}+4\st{H_2}{U}+12\tt{Q}{U}-2\tt{Q}{H_2}
+8\st{H_2}{Q}]$
\\ \\ \hline
\end{tabular}
\end{center}
\newpage
\begin{center}
\begin{tabular}{cccccc}\hline
\multicolumn{6}{c}{\bf TABLE III} \\ \hline
\multicolumn{6}{c}{$m_t=175 \: , \:  A_o=400 \: , \: m_o=300 \: , \:
m_{1/2}=200 \: , \: \mu(M_Z) >0$}
\\
$tan\beta$         &25       &20      &15       &10       &2
\\ \hline \\
$M_{GUT}$          & 2.632   &2.633   &2.633    &2.632    &2.515    \\
$(10^{16} GeV)$    &         &        &         &         &         \\
$\alpha_{GUT} $    &.04161    &.04161   &.04161    &.04161    &.04134   \\
$\alpha_{em}^{-1}$ &127.9    &127.9   &127.9    &127.9    &127.9    \\
$sin^{2}{\theta_W}$&.2311    &.2311   &.2311    &.2311    & .2311   \\
$\alpha_{3}$       &.13128   &.13129  &.13129   &.13126   &.12909   \\ \hline
\\
$M_t$              &177.0   &177.0    &177.0    &177.0    &176.8
\\ \hline   \\
$\tilde{g}$        &495.6   &495.8    &495.9    &495.9   &492.4     \\ \\
${\tilde{\chi}}_{1}^{o}$   &77.1      &77.0    &76.8    &76.4   &74.7    \\
${\tilde{\chi}}_{2}^{o}$   &139.1     &138.9   &138.5   &137.6  &136.0   \\
${\tilde{\chi}}_{3}^{o}$   &350.5     &352.3   &354.7   &359.1  &487.8   \\
${\tilde{\chi}}_{4}^{o}$   &-337.2    &-338.6  &-340.3  &-343.2 &-467.7  \\
\\
${\tilde{\chi}}_{1} ^{c}$  &352.3     &354.0   &356.0   &359.9  &484.5  \\
${\tilde{\chi}}_{2} ^{c}$  &138.8     &138.6   &138.1   &137.0  &134.7
\\  \hline \\
${\tilde{t}}_1$,${\tilde{t}}_2$  &527.3,315.4   &531.9,315.2   &535.9,314.4
				 &539.4,312.3   &543.9,285.3   \\
${\tilde{b}}_1$,${\tilde{b}}_2$  &506.2,441.0   &514.2,451.4   &520.9,459.4
				 &525.8,465.0   &525.9,462.4   \\ \\
${\tilde{\tau}}_1$,${\tilde{\tau}}_2$
				 &328.0,266.5   & 330.4,281.0  &331.7,292.9
				 &331.9,302.1   &329.4,309.2       \\
${\tilde{\nu}}_{\tau}$           &306.4         &311.5         &315.6
				 &318.6         &323.4       \\ \hline \\
${\tilde{u}}_{1,2}$,${\tilde{u}}_{1,2}^{c}$
				 &537.8,528.9  &537.8,528.9    &537.8,528.9
				 &537.7,528.8  &535.4,525.8      \\
${\tilde{d}}_{1,2}$,${\tilde{d}}_{1,2}^{c}$
				 &543.3,529.6  &543.3,529.6    &543.3,529.6
				 &543.2,529.5  &538.6,525.8      \\ \\
${\tilde{e}}_{1,2}$,${\tilde{e}}_{1,2}^{c}$
				 &330.1,311.7  &330.1,311.7   &330.0,311.7
				 &330.0,311.6  &328.8,310.3       \\
${\tilde{\nu}}_{1,2} $
				 &321.0        &321.0         &320.9
				 &321.0        &323.5      \\ \hline \\
$A$                &630.7        &612.0        &588.1   &560.1    &669.0  \\
$h_o$,$H_o$        &114.1,630.6  &114.2,611.9  &114.2,588.0
		   &113.8,560.2  &92.1,672.8                              \\
$H^{\pm}$          &635.4        &616.9        &593.1   &565.4     &673.5
\\ \hline
\end{tabular}
\end{center}

\newpage
\begin{center}
\begin{tabular}{cccccc}\hline
\multicolumn{6}{c}{\bf TABLE IV} \\ \hline
\multicolumn{6}{c}{$m_t=175 \: , \:  A_o=400 \: , \: m_o=300 \: , \:
tan\beta = 10 \: , \: \mu(M_Z) >0$}
\\
$m_{1/2}$         &700       &500     &300       &100       &75
\\ \hline \\
$M_{GUT}$          & 1.705   &1.915   &2.290    &3.145    &3.182    \\
$(10^{16} GeV)$    &         &        &         &         &         \\
$\alpha_{GUT} $    &.04030    &.04066   &.04120    &.04215    &.04222   \\
$\alpha_{em}^{-1}$ &127.9    &127.9   &127.9    &127.9    &127.9    \\
$sin^{2}{\theta_W}$&.2311    &.2311   &.2311    &.2311    & .2311   \\
$\alpha_{3}$       &.12669   &.12795  &.12983   &.13228   &.13110   \\ \hline
\\
$M_t$              &177.0   &177.0    &177.0    &177.0    &177.0
\\ \hline   \\
$\tilde{g}$        &1552.7  &1140.1   &714.6    &267.8    &207.1     \\ \\
${\tilde{\chi}}_{1}^{o}$   &291.5     &204.3   &118.6   &34.7   &24.4    \\
${\tilde{\chi}}_{2}^{o}$   &527.9     &371.3   &215.1   &62.2   &44.5   \\
${\tilde{\chi}}_{3}^{o}$   &752.0     &612.0   &450.5   &259.5  &234.7   \\
${\tilde{\chi}}_{4}^{o}$   &-733.9    &-595.8  &-434.8  &-243.8 &-219.5  \\
\\
${\tilde{\chi}}_{1} ^{c}$  &751.4     &611.6   &450.6   &261.9  &237.9  \\
${\tilde{\chi}}_{2} ^{c}$  &527.7     &371.1   &214.7   &60.7   &42.5
\\  \hline \\
${\tilde{t}}_1$,${\tilde{t}}_2$  &1347.6,1077.8 &1022.5,780.2   &698.2,470.9
				 &388.3,167.5   &353.0,144.7   \\
${\tilde{b}}_1$,${\tilde{b}}_2$  &1380.8,1311.8 &1039.9,975.9   &695.6,634.7
				 &370.2,305.3   &337.8,270.6   \\ \\
${\tilde{\tau}}_1$,${\tilde{\tau}}_2$
				 &548.5,390.1   &446.6,346.7    &362.5,313.7
				 &311.9,294.8   &308.8,293.8       \\
${\tilde{\nu}}_{\tau}$           &542.2         &438.4          &351.2
				 &296.9         &293.6       \\ \hline \\
${\tilde{u}}_{1,2}$,${\tilde{u}}_{1,2}^{c}$
			       &1444.7,1394.9  &1083.2,1049.7 &718.1,701.0
			       &372.1,371.1    &337.6,338.4      \\
${\tilde{d}}_{1,2}$,${\tilde{d}}_{1,2}^{c}$
			       &1446.6,1389.7  &1085.9,1046.7 &722.1,700.3
			       &380.0,373.4    &346.3,341.1      \\ \\
${\tilde{e}}_{1,2}$,${\tilde{e}}_{1,2}^{c}$
			       &550.0,399.3  &447.3,355.5   &361.8,322.8
			       &309.1,304.7  &305.9,303.7       \\
${\tilde{\nu}}_{1,2} $
				 &544.9        &440.9         &353.7
				 &299.3        &295.9      \\ \hline \\
$A$                &1341.8       &1036.5       &718.7   &414.6    &383.9  \\
$h_o$,$H_o$        &119.4,1341.9  &118.1,1036.6  &115.9,718.8
		   &108.7,414.7  &105.9,384.0                              \\
$H^{\pm}$          &1344.0       &1039.3    &722.8   &421.9     &391.8
\\ \hline
\end{tabular}
\end{center}

\newpage
\begin{center}
\begin{tabular}{cccccc}\hline
\multicolumn{6}{c}{\bf TABLE V} \\ \hline
\multicolumn{6}{c}{$m_t=175 \: , \:  tan\beta = 10 \: , \: m_o=300 \: , \:
m_{1/2}= 200 \: , \: \mu(M_Z) >0$}
\\
$Ao =$         &800       &-800     &300       &-300       &0
\\ \hline \\
$M_{GUT}$          & 2.574   &2.657   &2.645    &2.690    &2.676    \\
$(10^{16} GeV)$    &         &        &         &         &         \\
$\alpha_{GUT} $    &.04155    &.04165  &.04163    &.04169    &.04167   \\
$\alpha_{em}^{-1}$ &127.9     &127.9    &127.9    &127.9    &127.9    \\
$sin^{2}{\theta_W}$&.2311     &.2311    &.2311    &.2311    & .2311   \\
$\alpha_{3}$       &.13031    &.13183   &.13148   &.13235   &.13206  \\ \hline
\\
$M_t$              &177.0   &177.0    &177.0    &177.0    &177.0
\\ \hline   \\
$\tilde{g}$        &493.0  &501.1   &496.7    &500.3    &498.6     \\ \\
${\tilde{\chi}}_{1}^{o}$   &77.3      &75.9    &76.0    &73.7    &74.6    \\
${\tilde{\chi}}_{2}^{o}$   &141.8     &134.0   &136.1   &127.1   &130.6   \\
${\tilde{\chi}}_{3}^{o}$   &457.4     &305.2   &338.5   &271.3   &291.2   \\
${\tilde{\chi}}_{4}^{o}$   &-447.5    &-282.1  &-320.5  &-240.5  &-265.7  \\
\\
${\tilde{\chi}}_{1} ^{c}$  &458.4     &305.6   &339.2   &271.4  &291.5  \\
${\tilde{\chi}}_{2} ^{c}$  &141.6     &132.7   &135.3   &124.6  &128.9
\\  \hline \\
${\tilde{t}}_1$,${\tilde{t}}_2$  &522.6,182.5   &514.9,413.8   &541.6,332.8
				 &538.0,405.5   &543.4,378.5   \\
${\tilde{b}}_1$,${\tilde{b}}_2$  &520.8,433.0   &529.8,484.2   &526.8,470.9
				 &530.5,490.2   &529.2,484.1   \\ \\
${\tilde{\tau}}_1$,${\tilde{\tau}}_2$
				 &331.9,294.0   &328.0,302.5    &331.7,303.5
				 &330.1,306.8   &330.9,306.3       \\
${\tilde{\nu}}_{\tau}$           &316.2         &317.7          &319.0
				 &319.5         &319.6       \\ \hline \\
${\tilde{u}}_{1,2}$,${\tilde{u}}_{1,2}^{c}$
			       &536.1,527.2  &540.9,532.0 &538.2,529.2
			       &540.4,531.6  &539.4,530.5      \\
${\tilde{d}}_{1,2}$,${\tilde{d}}_{1,2}^{c}$
			       &541.6,527.8    &546.3,532.6   &543.6,529.9
			       &545.8,532.2    &544.8,531.2      \\ \\
${\tilde{e}}_{1,2}$,${\tilde{e}}_{1,2}^{c}$
			       &330.1,311.5  &330.4,311.6   &330.0,311.7
			       &330.1,311.7  &330.0,311.7       \\
${\tilde{\nu}}_{1,2} $
				 &321.1        &321.4         &321.0
				 &321.1        &321.0      \\ \hline \\
$A$                &661.3         &449.1        &537.2   &447.5    &480.5 \\
$h_o$,$H_o$        &119.6,660.8   &105.7,449.5  &112.7,537.4
		   &108.2,447.8   &110.1,480.8                          \\
$H^{\pm}$          &665.8         &455.7      &542.7     &454.1    &486.6
\\ \hline
\end{tabular}
\end{center}

\newpage
\begin{center}
\begin{tabular}{cccccc}\hline
\multicolumn{6}{c}{\bf TABLE VI} \\ \hline
\multicolumn{6}{c}{$m_t=175 \: , \:  tan\beta = 10 \: , \: Ao =400 \: , \:
m_{1/2}= 200 \: , \: \mu(M_Z) >0$}
\\
$ m_o$         &800       &600     &400       &200       &100
\\ \hline \\
$M_{GUT}$          & 2.687   &2.681   &2.660    &2.581    &2.482    \\
$(10^{16} GeV)$    &         &        &         &         &         \\
$\alpha_{GUT} $    &.04125   &.04138  &.04153    &.04169    &.04176   \\
$\alpha_{em}^{-1}$ &127.9     &127.9    &127.9    &127.9    &127.9    \\
$sin^{2}{\theta_W}$&.2311     &.2311    &.2311    &.2311    & .2311   \\
$\alpha_{3}$       &.13076    &.13087   &.13110   &.13142   &.13146 \\ \hline
\\
$M_t$              &177.0   &177.0    &177.0    &177.0    &177.0
\\ \hline   \\
$\tilde{g}$        &510.9  &504.7   &498.4    &494.2    &494.3     \\ \\
${\tilde{\chi}}_{1}^{o}$   &77.4      &77.1    &76.6    &76.2    &76.0    \\
${\tilde{\chi}}_{2}^{o}$   &137.2     &137.7   &137.7   &137.6   &137.5   \\
${\tilde{\chi}}_{3}^{o}$   &340.1     &351.6   &357.3   &360.6   &361.5   \\
${\tilde{\chi}}_{4}^{o}$   &-322.1    &-334.9  &-341.2  &-344.9  &-345.9  \\
\\
${\tilde{\chi}}_{1} ^{c}$  &340.6     &352.2   &358.0   &361.4  &362.3  \\
${\tilde{\chi}}_{2} ^{c}$  &136.4     &137.0   &137.0   &137.0  &137.0
\\  \hline \\
${\tilde{t}}_1$,${\tilde{t}}_2$  &762.2,522.9   &652.9,432.2   &569.7,349.3
				 &517.5,281.9   &504.0,260.9   \\
${\tilde{b}}_1$,${\tilde{b}}_2$  &879.9,734.4   &719.6,611.1   &581.4,507.1
				 &482.8,431.9   &454.6,410.1   \\ \\
${\tilde{\tau}}_1$,${\tilde{\tau}}_2$
				 &805.2,793.5   &611.8,596.2    &422.6,399.7
				 &247.5,205.7   &178.6,113.9       \\
${\tilde{\nu}}_{\tau}$           &800.1         &604.9          &412.3
				 &229.2         &152.3       \\ \hline \\
${\tilde{u}}_{1,2}$,${\tilde{u}}_{1,2}^{c}$
			       &890.2,887.3  &730.7,725.6 &593.2,585.5
			       &494.8,484.6  &466.5,455.2      \\
${\tilde{d}}_{1,2}$,${\tilde{d}}_{1,2}^{c}$
			       &893.5,888.2    &734.6,726.3   &598.1,586.2
			       &500.7,485.2    &472.8,455.8      \\ \\
${\tilde{e}}_{1,2}$,${\tilde{e}}_{1,2}^{c}$
			       &807.6,803.0  &612.8,604.9   &421.9,408.4
			       &244.0,217.6  &173.0,132.2       \\
${\tilde{\nu}}_{1,2} $
				 &804.1        &608.1         &414.9
				 &231.6        &155.0      \\ \hline \\
$A$                &910.8         &756.5        &618.1   &514.5    &485.0  \\
$h_o$,$H_o$        &112.3,911.0   &112.4,756.7  &113.2,618.2
		   &114.4,514.6   &114.8,485.0                          \\
$H^{\pm}$          &914.1         &760.4      &622.9     &520.3    &491.1
\\ \hline
\end{tabular}
\end{center}

\newpage
\begin{center}
\begin{tabular}{cccccc}\hline
\multicolumn{6}{c}{\bf TABLE VII} \\ \hline
\multicolumn{6}{c}{$tan\beta = 20 \: , \: Ao =200 \: , \: m_o = 400 \: , \:
m_{1/2}= 300 \: , \: \mu(M_Z) >0$}
\\
$ m_t $       &175       &180     &185       &190       &195
\\ \hline \\
$M_{GUT}$          & 2.337   &2.414   &2.495    &2.581    &2.670    \\
$(10^{16} GeV)$    &         &        &         &         &         \\
$\alpha_{GUT} $    &.04117   &.04118  &.04119    &.04120    &.04120   \\
$\alpha_{em}^{-1}$ &127.9     &127.9    &127.9    &127.9    &127.9    \\
$sin^{2}{\theta_W}$&.23110     &.23094    &.23078    &.23061    & .23044  \\
$\alpha_{3}$       &.13005    &.13058   &.13114   &.13171   &.13226 \\ \hline
\\
$M_t$              &177.0   &181.7    &186.4    &191.1    &195.8
\\ \hline   \\
$\tilde{g}$        &717.3  &719.0   &720.7    &722.5    &724.4     \\ \\
${\tilde{\chi}}_{1}^{o}$   &118.9     &119.0   &119.1   &119.2   &119.3    \\
${\tilde{\chi}}_{2}^{o}$   &213.2     &215.0   &216.4   &217.5   &218.5   \\
${\tilde{\chi}}_{3}^{o}$   &402.0     &422.0   &441.3   &459.9   &477.9   \\
${\tilde{\chi}}_{4}^{o}$   &-384.0    &-406.2  &-427.2  &-447.2  &-466.3  \\
\\
${\tilde{\chi}}_{1} ^{c}$  &402.6     &422.7   &442.1   &460.8  &478.8  \\
${\tilde{\chi}}_{2} ^{c}$  &212.8     &214.7   &216.1   &217.3  &218.3
\\  \hline \\
${\tilde{t}}_1$,${\tilde{t}}_2$  &720.3,526.9   &717.7,522.9   &715.2,519.5
				 &712.9,516.8   &710.7,515.1   \\
${\tilde{b}}_1$,${\tilde{b}}_2$  &727.0,665.1   &728.9,663.1   &730.8,661.3
				 &732.7,659.8   &734.6,658.7   \\ \\
${\tilde{\tau}}_1$,${\tilde{\tau}}_2$
				 &444.4,392.8   &444.9,392.2    &445.5,391.6
				 &446.0,391.1   &446.5,390.5       \\
${\tilde{\nu}}_{\tau}$           &432.2         &432.3          &432.4
				 &432.5         &432.5       \\ \hline \\
${\tilde{u}}_{1,2}$,${\tilde{u}}_{1,2}^{c}$
			       &761.3,745.7  &762.7,747.2 &764.3,748.7
			       &765.9,750.3  &767.5,752.0      \\
${\tilde{d}}_{1,2}$,${\tilde{d}}_{1,2}^{c}$
			       &765.1,745.2    &766.6,746.6   &768.1,748.2
			       &769.7,749.8    &771.3,751.5      \\ \\
${\tilde{e}}_{1,2}$,${\tilde{e}}_{1,2}^{c}$
			       &447.3,417.0  &447.3,417.0   &447.4,417.0
			       &447.4,417.1  &447.5,417.1       \\
${\tilde{\nu}}_{1,2} $
				 &440.7        &440.8         &440.8
				 &440.9        &441.0      \\ \hline \\
$A$                &795.6         &820.6        &843.0   &862.8    &879.8  \\
$h_o$,$H_o$        &115.0,795.6   &116.9,820.6  &118.8,843.0
		   &120.7,862.7   &122.6,879.7                          \\
$H^{\pm}$          &799.3         &824.1      &846.4     &866.1    &883.1
\\ \hline
\end{tabular}
\end{center}

\newpage
\begin{center}
\begin{tabular}{cccc}\hline
\multicolumn{4}{c}{\bf TABLE VIII} \\ \hline
\multicolumn{4}{c}{$m_t=175$, $tan\beta=10$, $A_o=250$, $m_o=200$,
$m_{1/2}=150$, $\mu(M_Z) >0$} \\ \hline
	  & Case [a]    &  Case [b]        & Case [c]             \\
	  & 1-loop      &  2-loop          & Complete 2-loop        \\
\hline
$M_{GUT}$          & 2.1881    &2.8876        &2.8766           \\
$(10^{16} GeV)$    &                  &             &         \\
$\alpha_{GUT} $    &.04127     &.04201       &.04202          \\
$\alpha_{em}^{-1}$ &127.9             &127.9        &127.9    \\
$sin^{2}{\theta_W}$&.23105            &.23110       &.23110  \\
$\alpha_{3}$       &.11767    &.13284        &.13289       \\ \hline
\\
$M_t$              &181.0             &177.0         &177.0
\\ \hline   \\
$\tilde{g}$        &398.4      &381.4       &382.6            \\ \\
${\tilde{\chi}}_{1}^{o}$   &59.2  &55.0   &54.4           \\
${\tilde{\chi}}_{2}^{o}$   &109.0  &98.3  &96.7           \\
${\tilde{\chi}}_{3}^{o}$   &302.8     &304.1  &279.6       \\
${\tilde{\chi}}_{4}^{o}$   &-284.1    &-287.7   &-260.1       \\
\\
${\tilde{\chi}}_{1} ^{c}$  &304.0        &305.5   &280.9       \\
${\tilde{\chi}}_{2} ^{c}$  &108.0   &97.4       &95.3
\\  \hline \\
${\tilde{t}}_1$,${\tilde{t}}_2$  &443.6,247.0   &442.0,235.2    &440.2,234.7
  \\
${\tilde{b}}_1$,${\tilde{b}}_2$  &401.7,357.2  &395.4,352.8   &394.9,352.6
   \\ \\
${\tilde{\tau}}_1$,${\tilde{\tau}}_2$
				 &235.3,203.6 &232.2,201.6  &231.3,202.7
   \\
${\tilde{\nu}}_{\tau}$           &216.6       &212.6     &212.5
   \\ \hline \\
${\tilde{u}}_{1,2}$,${\tilde{u}}_{1,2}^{c}$
			       &410.5,401.4   &400.1,394.1  &400.1,394.1
  \\
${\tilde{d}}_{1,2}$,${\tilde{d}}_{1,2}^{c}$
			       &417.8,402.4 &407.5,395.7   &407.5,395.7
  \\ \\
${\tilde{e}}_{1,2}$,${\tilde{e}}_{1,2}^{c}$
			       &231.6,212.6  &227.7,211.6  &227.5,211.8
  \\
${\tilde{\nu}}_{1,2} $
				 &218.1      &214.2     &214.1
  \\ \hline \\
$A$                       &412.1       &421.0        &394.5           \\
$h_o$,$H_o$               &113.0,412.2 &110.7,421.1  &110.5,394.7     \\
$H^{\pm}$                 &419.5       &428.1        &402.1
\\ \hline
\end{tabular}
\end{center}

\end{document}